\documentclass[%
 reprint,
 amsmath,amssymb,
 aps,prd,longbibliography,
floatfix,
]{revtex4-1}

\usepackage{graphicx}% Include figure files
\usepackage{dcolumn}% Align table columns on decimal point
\usepackage{bm}% bold math
\usepackage{lipsum}

\newcommand{\mn}{{Monthly Notices of the Royal Astronomical Society}}
\newcommand{\mnras}{\mn}

\newcommand{\apjl}{{Astrophysical Journal Letters}}

\newcommand{\aap}{{Astronomy \& Astrophysics}}
\newcommand{\araa}{{Annual Review of Astronomy \& Astrophysics}}
\newcommand{\procspie}{Proc. SPIE}
\begin{document}

\preprint{APS/123-QED}

\title{Vacuum birefringence and the x-ray polarization from black-hole accretion disks}% Force line breaks with \\

\author{Ilaria Caiazzo}%
\email{ilariacaiazzo@phas.ubc.ca}%
\author{Jeremy Heyl}%
\affiliation{%
Department of Physics and Astronomy, University of British
  Columbia, 6224 Agricultural Road, Vancouver, British Columbia V6T 1Z1, Canada \\
}%

\date{\today}% It is always \today, today,

\begin{abstract}
In the next decade, x-ray polarimetry will open a new window on the high-energy Universe, as several missions that include an x-ray polarimeter are currently under development. Observations of the polarization of x-rays coming from the accretion disks of stellar-mass and supermassive black holes are among the new polarimeters' major objectives. In this paper, we show that these observations can be affected by the quantum electrodynamic (QED) effect of vacuum birefringence: after an x-ray photon is emitted from the accretion disk, its polarization changes as the photon travels through the accretion disk's magnetosphere, as a result of the vacuum becoming birefringent in presence of a magnetic field. We show that this effect can be important for black holes in the energy band of the upcoming polarimeters, and has to be taken into account in a complete model of the x-ray polarization that we expect to detect from black-hole accretion disks, both for stellar mass and for supermassive black holes. We find that, for a chaotic magnetic field in the disk, QED can significantly decrease the linear polarization fraction of edge-on photons, depending on the spin of the hole and on the strength of the magnetic field. This effect can provide, for the first time, a direct way to probe the magnetic field strength close to the innermost stable orbit of black-hole accretion disks and to study the role of magnetic fields in astrophysical accretion in general.  
%\begin{description}
%\end{description}
\end{abstract}

\pacs{Valid PACS appear here}% PACS, the Physics and Astronomy
\maketitle

%\tableofcontents

\section{\label{sec:level1}Introduction}
In quantum electrodynamics (QED), the vacuum is expected to be birefringent in the presence of a magnetic field. This effect is extremely small for the magnetic fields that are obtainable here on Earth, and therefore, this prediction, which was one of the first predictions of QED \cite{1936...Heisen...Euler,1936...Weisskopf,PhysRev.82.664}, has not been tested until very recently. Astrophysical magnetic fields, on the other hand, are much more intense, and objects like neutron stars and black holes, which are characterized by magnetic fields that can be as large as $10^{15}$~G, are ideal laboratories to test this prediction. Recently, the detection of the optical polarization of a radio-quiet neutron star, RX~J1856.5-3754, has provided a hint that vacuum birefringence actually exists \cite{2017MNRAS.465..492M}.

The effect of the vacuum birefringence on the photon's polarization is much stronger in the x-rays \cite{2000MNRAS.311..555H}, and, indeed, several studies show how the observation of x-ray polarization from neutron stars can, on one hand, test QED and, on the other hand, probe the magnetosphere of these objects \cite{2002PhRvD..66b3002H,2014MNRAS.438.1686T}. Black-hole accretion disks are expected to generate a weaker magnetic field, so one could think that the effect of vacuum birefringence on x-ray polarization would be small. However, how strongly the birefringence affects the polarization of the photon traveling in the magnetized vacuum does not depend only on the strength of the magnetic field itself, but also on for how long the photon travels in the strong magnetic field (see Sec.~\ref{sec:QED}). In Sec.~\ref{sec:NT}, we estimate the photon energy at which QED becomes important as a function of the black hole spin and for a magnetic field strength that is the minimum needed for accretion to occur in a $\alpha$ model. We show that, for fast spinning black holes, QED becomes important at the high end of the upcoming polarimeters' range, around 10~keV. If the magnetic field was in fact stronger than the minimum required by the model, the effect would be important at lower energies as well. The estimate in Sec.~\ref{sec:NT}, however, does not take into account the structure of the magnetic field and the angular momentum of the photon. In order to get a better sense of how QED would affect the final polarization of x-ray photons, we calculate the effect of a partially ordered magnetic field on x-rays traveling near the disk plane (see Sec.~\ref{sec:case}).

A detailed calculation of the effect of QED on x-ray polarization from black holes is crucial as several missions with x-ray polarimeters are currently under development. In the soft range (1 to 10 keV), the NASA mission IXPE \cite{2016SPIE.9905E..17W} has already been selected in the current SMEX cycle, XIPE is among the 3 ESA M4 candidates \cite{2016SPIE.9905E..15S}, and CAS/CNSA mission eXTP is currently in its phase zero \cite{2016SPIE.9905E..1QZ}. In the sub-keV range, a narrow band (250 eV) polarimeter, LAMP, and a broad band (0.2-0.8 keV) rocket-based polarimeter, REDSox, are currently being designed \cite{2015SPIE.9601E..0IS,SPIE_REDSoX}. As for hard x-rays, the balloon-borne X-Calibur \cite{2014JAI.....340008B}, with an energy range of 20 to 80 keV, and POGO+ \cite{2018MNRAS.tmpL..30C}, 18 to 160 keV, have already observed the Crab and Cygnus X-1.

Magnetic fields have an important role in the physics of black holes, as they are thought to lead to the formation of relativistic jets \cite{1977MNRAS.179..433B,2011MNRAS.418L..79T} and to be the source of the shear stresses in accretion disks \cite{1973A&A....24..337S,1991ApJ...376..214B}. However, there have been very few observations capable of probing the strength of magnetic fields in accretion disks, especially close to the black hole. Recently, observations of the radio polarization from Sagittarius A*, the supermassive black hole at the center of our own Galaxy, have found evidence for partially ordered magnetic fields near the event horizon \cite{2015Sci...350.1242J}. The work of \citet{2006Natur.441..953M,2008ApJ...680.1359M,2016ApJ...821L...9M} on spectra from stellar mass black hole binaries, has shown that the observed wind, produced by the inner disk as close as 850 $GM/c^2$ to the central engine, must be powered by magnetic processes, most likely by pressure generated by magnetic shear stresses internal to the disk, and can provide an indication on the strength of the magnetic field in that region. The observation of the x-ray polarization from black-hole accretion disks could be, if properly modeled including QED, the first probe of black-hole magnetic fields close to the innermost stable orbit of the disk, and a  direct way to study the role of magnetic fields in astrophysical accretion generally.

\section{Vacuum Birefringence}
\label{sec:QED}

In classical electrodynamics, Maxwell's equations are linear in the fields, and there is no interaction of light with light. In QED, the vacuum current of a charged Dirac field implies an addition to the action integral of the electromagnetic field, as photons can interact with the field via the production of virtual electron-positron pairs.
This addition can be written as a function of Lorentz- and gauge-invariant quantities \cite{PhysRev.82.664}:
\begin{subequations}
\begin{equation}
F = \frac{1}{4} F_{\mu\nu}F^{\mu\nu} = \frac{1}{2} (\bm{B}-\bm{E})^2
\end{equation}
\begin{equation}
G = \frac{1}{4} F_{\mu\nu} F^*_{\mu\nu} = \bm{E}\cdot\bm{B}
\end{equation}
\begin{equation}
X^2 = 2(F + i G) = (\bm{B} + i \bm{E})^2
\end{equation}
\end{subequations}
where $\bm{E}$ and $\bm{B}$ are constant electric and magnetic field strengths and $F_{\mu\nu}$ is the electromagnetic tensor.
The full Lagrangian density of constant electromagnetic fields becomes
\begin{widetext}
\begin{equation}
\label{eq:lagrangian}
\mathcal{L} = - F -  \frac{1}{8\pi}\int_0^\infty ds s^{-3} \exp{(-m^2s)} \left\{ (es)^2 G\frac{\Re [\cosh (esX)]}{\Im [\cosh (esX)]} - 1 - \frac{2}{3}(es)^2 F \right\} \; .
\end{equation}
\end{widetext}
In Eq.~(\ref{eq:lagrangian}), the term $-F$ is the classical Lagrangian density of electrodynamics and the second term is the addition due to the presence of a charged Dirac field in QED.

Historically, \citet{1936...Heisen...Euler} and \citet{1936...Weisskopf} independently derived an effective Lagrangian for weak fields using electron-hole theory. \citet{PhysRev.82.664} later derived Eq.~(\ref{eq:lagrangian}) using QED, which gives the same result for the effective Lagrangian in the weak field limit:
\begin{widetext}
\begin{equation}
\label{eq:lageff}
  \mathcal{L}_{\rm{eff}} \simeq \frac{1}{2}(\mathbf{E}^2 - \mathbf{B}^2) + \frac{2\alpha_\mathrm{QED}^2}{45} \frac{(\hbar/mc)^3}{mc^2}[(\mathbf{E}^2 -\mathbf{B}^2)^2 + 7(\mathbf{E}\cdot\mathbf{B})^2]
\end{equation}
\end{widetext}

The non-linear interaction of the electromagnetic field implies that, if you add a photon field as a perturbation to the constant electromagnetic field, the speed at which light travels through the magnetized vacuum depends on its direction, polarization and on the local strength of the magnetic field. In other words, the vacuum in presence of a magnetic field acquires an index of refraction $n$ different from unity.  From the effective Lagrangian in Eq.~(\ref{eq:lageff}), \citet{1997JPhA...30.6485H} calculated the difference between the index of refraction in the direction perpendicular to the magnetic field and the one in the parallel direction. In the weak field limit ($B \ll 0.5 B_{\mbox{\tiny{QED}}}$):
\begin{equation}
  \label{eq:deltan}
  n_{\parallel} - n_{\perp} = \frac{\alpha_{\mbox{\tiny{QED}}}}{30 \pi} \left(\frac{B}{B_{\mbox{\tiny{QED}}}} \right)^2\sin^2\theta
\end{equation}
where $\theta$ is the angle between the direction of the wave number of the photon and the external field,
$\alpha_{\mbox{\tiny{QED}}}$ is the fine structure constant and
\begin{equation}
  B_{\mbox{\tiny{QED}}} = \frac{m_e^2 c^3} {\hbar e} \simeq 4.4 \times 10^{13} \mbox{ G} \, .
\end{equation}

  In order to describe the evolution of the polarization of a single photon in the magnetized vacuum, we employ the Poincar\'e sphere formalism: polarization is described by a unit vector
\begin{equation}
  \mathbf{s} = \frac{1}{S_0}(S_1,S_2,S_3) = \frac{1}{I}(Q,U,V).
\end{equation}
where $S_0$, $S_1$, $S_2$, $S_3$ (or $I$, $Q$, $U$, $V$) are the Stokes parameters and the polarization states are mapped on the surface of a sphere.
\citet{Kubo:81,Kubo:83} derived the equation of motion of the polarization direction on the Poincar\'e sphere as light travels through an inhomogeneous birefringent medium. They obtain the following expression for the evolution of the polarization of a wave:
\begin{equation}
  \label{eq:pol}
  \frac{\partial \mathbf{s}}{\partial x_3} = \hat{\bm{\Omega}} \times\mathbf{s} + (\hat{\mathbf{T}}\times\mathbf{s}) \times \mathbf{s}
\end{equation}
where $\hat{\bm{\Omega}}$ is the birefringent vector, $\hat{\mathbf{T}}$ is the dichroic vector and $x_3$ is the length of the photon path on the sphere.

In the case of vacuum in QED with an external magnetic field to one loop and weak electric field, $\hat{\mathbf{T}} = 0$ (there is no real pair production) and the amplitude of $\hat{\bm{\Omega}}$ is proportional to the difference between the indices of refraction for the two polarization states. From Eq.~(\ref{eq:deltan}),
\begin{equation}
  \label{eq:birivector}
  \hat{\Omega} = k_0(n_{\parallel} - n_{\perp}) = k_0\frac{\alpha_{\mbox{\tiny{QED}}}}{30 \pi} \left(\frac{B}{B_{\mbox{\tiny{QED}}}}\right)^2\sin^2\theta
\end{equation}
where $k_0 = 2 \pi \nu / c$, is the unperturbed wave number of the photon.

All the current models for the polarization of light emitted from a black hole \citep{2009ApJ...701.1175S,2010ApJ...712..908S,2011ApJ...731...75D,2016ApJ...819...48S} assume that the polarization vectors of the photons are simply  parallel transported along the photon paths as the photons travel through the black hole spacetime. This assumption may fail, because, as photons travel in a strong magnetic field, their polarization can change along the path due to vacuum birefringence.
Numerical integrations of Eq.~(\ref{eq:pol}) show that, for strong enough magnetic fields, the photon polarization modes are decoupled, and the direction of polarization follows the direction of the magnetic field \cite{2000MNRAS.311..555H}. This effect becomes important when
\begin{equation}
  \label{eq:h&s}
  \left| \bm{\hat{\Omega}} \left( \frac{1}{|\bm{\hat{\Omega}}|} \frac{\partial |\bm{\hat{\Omega}}|}{\partial x_3} \right)^{-1} \right| \gtrsim 0.5.
\end{equation}
From the same expression, we can find the distance from the source at which the polarization stops following the magnetic field and becomes frozen with respect to the radius curvature. This distance is called the polarization-limiting radius, or the adiabatic radius (see Sec.~\ref{sec:NT}). We want to calculate the polarization-limiting radius for black holes in order to understand whether QED is important for x-ray observations and, therefore, we need to estimate the strength of the magnetic field in the disk.

\section{Accretion disk model}
\label{sec:NT}

Black-hole accretion disks are rarefied; thus, angular momentum transfer due to molecular viscosity is inefficient and cannot lead to accretion \cite{1981ARA&A..19..137P}. In current theories of astrophysical accretion disks, magnetic fields and turbulence are expected to be the source of shear stresses. Here, in order to have an estimate of the strength of the magnetic field in the disk plane,  we set the magnetic field to be the minimum needed for accretion to occur. The relation between the tangential stresses between layers in the disks and the magnetic field can be written as \cite{1973A&A....24..337S}
\begin{equation}
  \label{eq:stress}
  t_{\hat{\phi}\hat{r}} = \rho c_s v_t + \frac{B^2}{4\pi} = \alpha P
\end{equation}
where $\rho$ is the mass density, $c_s$ is the speed of sound, $v_t$ is the turbulence velocity, $P$ is pressure and $t_{\hat{\phi}\hat{r}}$ is the shear stress as measured in a frame of reference moving with the gas. The last equality is called the $\alpha-$prescription, in which the efficiency of the angular momentum transfer is expressed with one parameter. Since turbulence in the disk is generated by the magnetic field, we expect the two terms to be of the same order. The minimum strength for the magnetic field to generate the shear stresses needed for accretion is then of the order $B\sim (4\pi\alpha P)^{1/2}$.

We model the accretion disk physics using the Novikov and Thorne (N\&T) model~\cite{1973blho.conf..343N}. The N\&T accretion disk model is the general relativistic generalization of the Shakura-Sunyaev model \cite{1973A&A....24..337S}, set in the Kerr spacetime surrounding a spinning black hole with spin parameter $a = J/(cM)$, ranging from $a=0$ (Schwarzschild black hole) to $a=GM/c^2$ (critical spin).

For simplicity, in order to split expressions into Newtonian limits times relativistic corrections, N\&T introduced the following functions (from now on we will use $c=G=1$), which are equal to one in the nonrelativistic limit:
\begin{subequations}
\label{eq:GR}
\begin{align}
  \mathcal{A} =& 1 + a_\star^2/r_\star^2 + 2a_\star^2/r_\star^3 \\ 
  \mathcal{B} =& 1 + a_\star/r_\star^{3/2} \\
  \mathcal{C} =& 1 - 3/r_\star + 2a_\star/r_\star^{3/2} \\
  \mathcal{D} =& 1 -2/r_\star +a_\star^2/r_\star^2 \\
  \mathcal{E} =& 1 + 4a_\star^2/r_\star^2 - 4a_\star^2/r_\star^3 + 3a_\star^4/r_\star^4 \\
  \mathcal{F} =& 1 - 2a_\star/r_\star^{3/2} + a_\star^2/r_\star^2 \\
  \mathcal{G} =& 1 - 2/r_\star + a_\star/r_\star^{3/2} \\
  \mathcal{N} =& 1 - 4 a_\star/r_\star^{3/2} + 3a_\star^2/r_\star^2
\end{align}
\end{subequations}
where $r_\star=r/M$ and $a_\star=a/M$. The last expression, $\mathcal{N}$, which is not from \citet{1973blho.conf..343N}, corresponds to the quantity called $C$ in \citet{1995ApJ...450..508R}.
In the N\&T accretion disk model, the disk lies in the equatorial plane ($\theta=\pi/2$) and matter rotates in quasi-circular orbits with angular velocity
\begin{equation}
  \omega = \frac{d\phi}{dt} = \sqrt{\frac{M}{r^3}}\frac{1}{\mathcal{B}}.
\end{equation}
\iffalse
From the Kerr metric near the equatorial plane ($\theta \sim \pi/2$),
\begin{align}
  &ds^2 = -\frac{r^2\Delta}{A}dt^2 + \frac{A}{r^2}(d\phi - \omega dt)^2 + \frac{r^2}{\Delta}dr^2 + dz^2 \\
  &\Delta \equiv r^2 - 2Mr + M^2a^2 = r^2\mathcal{D} \nonumber \\
  &A = r^4 + M^2r^2a^2+2M^3ra^2 = r^4 \mathcal{A} \nonumber \\
  &\omega \equiv 2Mar/A = (2Ma/r^3)\mathcal{A}^{-1} \nonumber 
\end{align}
where $z=r\cos\theta \sim r(\theta - \pi/2)$.
\fi
The inner edge of the disk is determined by the innermost stable circular orbit (or ISCO), which lies at the radius
\begin{align}
  &r_I = M\{3+Z_2 - [(3-Z_1)(3+Z_1+2Z_2)]^{1/2}\} \\
  &Z_1 \equiv 1 + (1-a_\star^2)^{1/3}[(1+a_\star)^{1/3} + (1-a_\star)^{1/3}] \nonumber \\
  &Z_2 \equiv (3a_\star^2 + Z_1^2)^{1/2} \nonumber
\end{align}
  
In order to calculate the pressure in the disk, we have to analyze the local vertical structure of the disk near the equatorial plane. The easiest way is to perform the calculations in the local orbiting frame at the center of the disk ($z=0$). In this inertial frame of reference, all that is needed are the following equations, in which the Newtonian value is multiplied by the relativistic corrections defined in Eqs.~(\ref{eq:GR}).
We will need, of course, the equation for hydrostatic equilibrium in general relativity. We use the correction to the N\&T equilibrium found by Riffert and Herold \cite{1995ApJ...450..508R}:
\begin{equation}
  \label{eq:hydro}
  \frac{dP}{dSec.igma} = - \omega^2 z \frac{\mathcal{B}^2\mathcal{N}}{\mathcal{C}}
\end{equation}
where $dSec.igma = \rho dz$. Since we are interested in the mid-plane, where by symmetry we expect the vertical density profile to reach a local maximum (or minimum), we consider $\rho$ to be approximately constant near the mid-plane. We then need an expression for how the energy is generated inside the disk. The viscous heating generated by friction between adjacent layers is given by \cite{1973blho.conf..343N}
\begin{equation}
  \label{eq:heating}
  \frac{dF}{dz} = \frac{3}{2}\omega \, t_{\hat{\phi}\hat{r}} \, \mathcal{C}^{-1}\mathcal{BD}
\end{equation}
where $F$ is the energy flux. 
We assume the energy transport to be radiative:
\begin{equation}
  \label{eq:transport}
  F = -\frac{1}{\kappa_{\mbox{\tiny{R}}}} \frac{dP_{\mbox{\tiny{rad}}}}{dSec.igma}
\end{equation}
where $\kappa_{\mbox{\tiny{R}}}$ is the Rosseland mean opacity.
For the equation of state to calculate the vertical structure, we 
assume that, in the central part of the disk, pressure is dominated by radiation. However, we allow a $z$ dependence:
\begin{equation}
 \label{eq:eos}
  P = \frac{1}{\chi(Sec.igma)}P_{\mbox{\tiny{rad}}} \; .
\end{equation}

From Eqs.~(\ref{eq:hydro}), (\ref{eq:transport}), and (\ref{eq:eos}) we
get:
\begin{align}
  - \kappa_R F &= \frac{d(\chi(Sec.igma) P)}{dSec.igma} = \frac{d\chi}{dSec.igma} P + \chi \frac{dP}{dSec.igma} \nonumber \\
  &= \frac{d\chi}{dSec.igma} P + \chi(-\omega^2 z) \frac{\mathcal{B}^2\mathcal{N}}{\mathcal{C}}
\end{align}
Thus, from Eqs.~(\ref{eq:heating}) and (\ref{eq:eos}):
\begin{widetext}
\begin{equation}
  \alpha P = \chi \frac{2 \omega}{3 \kappa_R}  \frac{\mathcal{B}\mathcal{N}}{\mathcal{D}} - \frac{2F}{3\omega} \frac{d\ln \kappa_R}{dz} \frac{\mathcal{C}}{\mathcal{BD}} + \frac{2 \omega z}{3 \kappa_R} \frac{d\chi}{dz}  \frac{\mathcal{B}\mathcal{N}}{\mathcal{D}} -  \frac{2}{3\kappa_R \omega} \frac{d}{dz} \left( \frac{d\chi}{dSec.igma} P \right)\frac{\mathcal{C}}{\mathcal{BD}}
\end{equation}
\end{widetext}
In the mid-plane this becomes:
\begin{align}
  \alpha P_c &= \chi_c \frac{2\omega}{3 \kappa_R}  \frac{\mathcal{B}\mathcal{N}}{\mathcal{D}}  - P_c \frac{2}{3 \kappa_R \omega} \frac{d}{dz} \left. \left(\frac{d\chi}{dSec.igma} \right) \right|_{z=0} \frac{\mathcal{C}}{\mathcal{BD}} \nonumber \\
 &\sim \chi_c \frac{2\omega}{3 \kappa_R}  \frac{\mathcal{B}\mathcal{N}}{\mathcal{D}}  - P_c \frac{2}{3 \kappa_R \omega} \left. \frac{\chi_c}{\rho_c h^2} \right|_{z=0} \frac{\mathcal{C}}{\mathcal{BD}} 
\end{align}
where $h$ is the typical scale height of the disk.
The second term is negative because $\chi$ decreases with $z$ and $Sec.igma$ and
it reaches its maximum at $z=0$, so its derivative at $z=0$ is less than 0.

Rewriting $\kappa_R \rho_c = 1/\lambda$ (mean free path), we have:
\begin{equation}
  P_c \frac{2}{3\kappa_R \omega}\frac{\chi_c}{\rho_c h^2} = \chi_c P_c \frac{2 \lambda^2}{\omega \lambda h^2}
\end{equation}
In this expression, $h^2/\lambda^2$ corresponds to the number of mean free paths
that a photon needs to perform a random walk out of the disk, while
$\lambda/c$ is the time for one mean free path. We can then rewrite this
expression in terms of the diffusion time:
\begin{equation}
  \chi_c P_c \frac{2 \lambda^2}{\omega \lambda h^2} = \chi_c P_c \frac{2}{\omega t_{\mbox{\tiny{diff}}}} = \chi_c P_c \frac{t_\mathrm{rot}}{\pi t_{\mbox{\tiny{diff}}}}
\end{equation}
where $t_\mathrm{rot}$ is the time needed by the disk to undergo a complete rotation and $t_{\mbox{\tiny{diff}}}$ is the diffusion time. Since $t_\mathrm{rot}\ll t_{\mbox{\tiny{diff}}}$, this term is much smaller than the first one. The relativistic corrections do not affect this result because the value of $\mathcal{C}/(\mathcal{BD})$ is less than one from the ISCO to infinity and it goes to one at infinity. We can then write the strength of the magnetic field in the mid-plane as:
\begin{equation}
  \label{eq:B}
  B^2 \sim 4\pi\alpha P_c \sim \chi_c \frac{2\omega}{3\kappa_R} \frac{\mathcal{B}\mathcal{N}}{\mathcal{D}} = \chi_c \frac{8\pi}{3\kappa_R} \sqrt{\frac{M}{r^3}} \frac{\mathcal{N}}{\mathcal{D}} \; .
\end{equation}
Since radiation dominates the pressure in the mid-plane of the disk, we can take $\chi_c \sim 1$. Moreover, it is safe to assume that in the innermost part of the disk the opacity is dominated by electron scattering:
\begin{equation}
  \kappa_R = \kappa_{es} = \frac{8 \pi}{3 m_p} \left( \frac{e^2}{m_e c^2} \right)^2 \frac{(1 + X)}{2}
\end{equation}
where $m_p$ and $m_e$ are the proton mass and the electron mass respectively and $X$ is the hydrogen mass fraction.
For a 10 M$_\odot$ black hole at the ISCO, $r=r_I$, we obtain
\begin{equation}
\label{eq:bfield}
	B^2 = (0.36 - 1.22 \times 10^8 \, \mathrm{G})^2 \left( \frac{M}{10 \, \mathrm{M}_\odot} \right)^{-1} \left(\frac{1+X}{2} \right)^{-1}
\end{equation}
where the first value is for $a_\star=0$ and the second is for $a_\star=0.999 $ (the value diverges for $a_\star=1$). This is a crude estimate of the minimum magnetic field strength needed to generate enough shear stresses for accretion to occur. Both global magneto-hydrodynamic (MHD) simulations \cite{2013ApJ...769..156S} and shearing box simulations \cite{2009ApJ...691...16H} show that, when moving away from the mid-plane, the magnetic pressure decreases toward the photosphere. However, our expression, Eq.~(\ref{eq:bfield}) with the radial scaling of Eq.~(\ref{eq:B}), reproduces the strength of the magnetic field at the photosphere obtained with shearing box simulations by \citet{2009ApJ...691...16H} for a 6.62~M$_\odot$ black hole at a radius of 30 $GM/c^2$. Likewise, the expressions in Eqs.~(\ref{eq:B}) and~(\ref{eq:bfield}) reproduce both the strength and the radial decrease of the magnetic field along the photosphere in Fig.~3 of \citet{2013ApJ...769..156S}, who performed a global MHD simulation for a 10~M$_\odot$ black hole. Regarding the estimates obtained by \citet{2016ApJ...821L...9M} for GRS 1915+105 at 850, 1,200, 3,000 and 30,000 $GM/c^2$, Eqs.~(\ref{eq:B}) and~(\ref{eq:bfield}) reproduces their minimum estimate at every radius, the one obtained by assuming MHD pressure, while it is two orders of magnitude less than their estimates obtained by assuming a magnetocentrifugal driven wind or an $\alpha$-model pressure.  For our purposes, we will then use the analytical expression found in Eq.~(\ref{eq:B}) for the minimum magnetic field strength at the photosphere.

From equations~(\ref{eq:birivector}) and~(\ref{eq:B}), we can estimate the amplitude of the birefringent vector in the vacuum just above the accretion disk. Reintroducing all the constants yields the magnitude of the birefringent vector:
\begin{equation}
	\label{eq:Omega}
	\hat{\Omega} = k_0 \Delta n = k_0 \frac{\hbar m_p}{15 \pi m_e^2c^2} \frac{1}{(1+X)} \sqrt{\frac{GM}{r^3}} \frac{\mathcal{N}}{\mathcal{D}} \sin^2 \theta
\end{equation}
And Eq.~(\ref{eq:h&s}) becomes
\begin{align}
  &\left| \bm{\hat{\Omega}} \left( \frac{1}{|\bm{\hat{\Omega}}|} \frac{\partial |\bm{\hat{\Omega}}|}{\partial x_3} \right)^{-1} \right| \simeq \hat{\Omega}(r) r \nonumber \\
  &\simeq k_0 \frac{\hbar m_p}{15 \pi m_e^2c^2} \frac{1}{(1+X)} \sqrt{\frac{GM}{r}} \frac{\mathcal{N}(r)}{\mathcal{D}(r)} \; .
\end{align}
Equating this expression to 0.5, we can calculate the polarization limiting radius to be
\begin{equation}
\frac{r_p c^2}{GM} = \left(\frac{2k_0\hbar m_p}{15\pi  m_e^2 c (1+X)} \frac{ \mathcal{N}(r_p)}{\mathcal{D}(r_p)} \right)^2
\end{equation}

\begin{figure}
\includegraphics[width=\columnwidth]{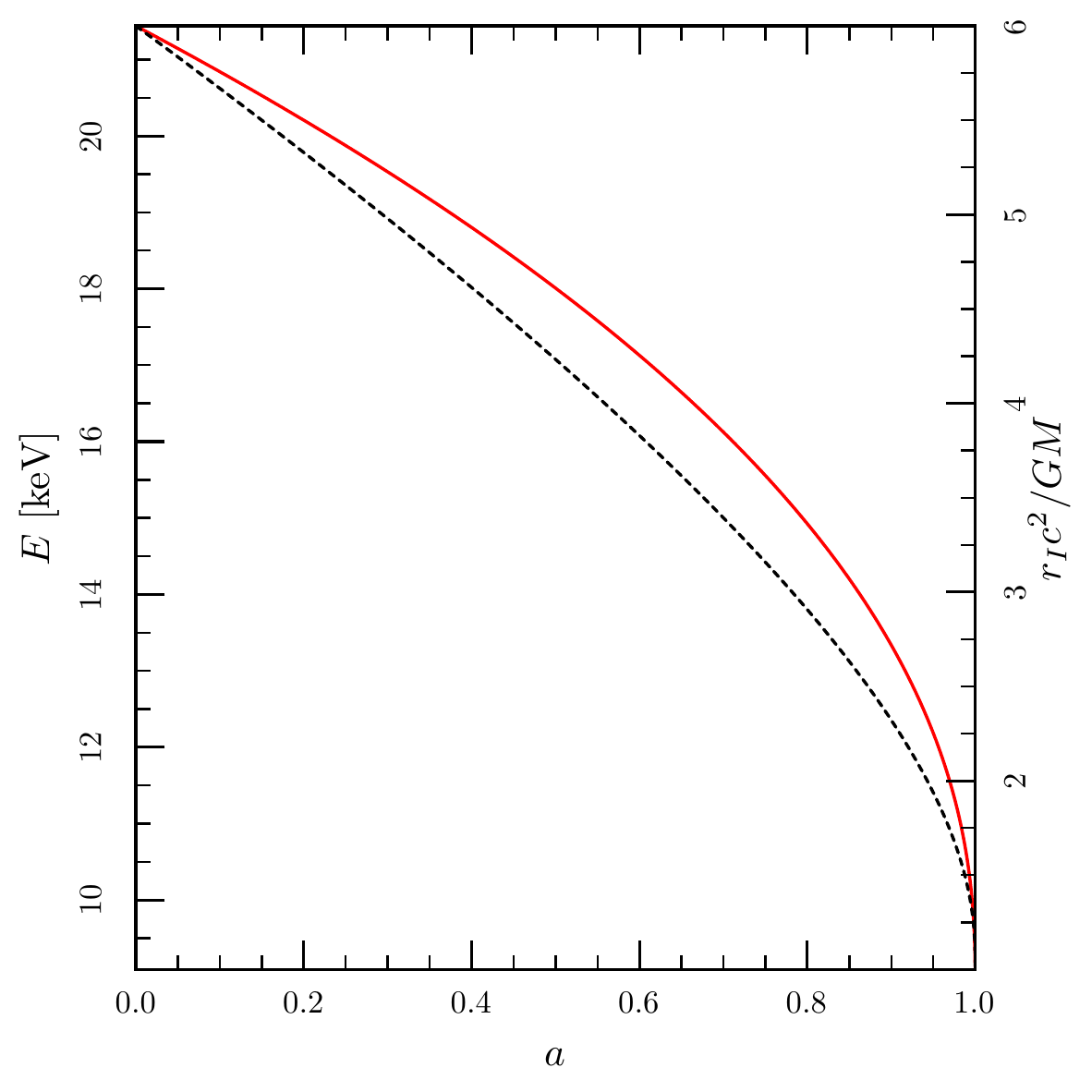}
\caption{The plot shows, on the left, $y$ axis, the energy at which $r_p = r_I$ (solid red line).  On the right, $y$ axis, the ISCO for a black hole as function of the spin parameter $a$ (dashed black line).}
\label{fig:pol_rad}
\end{figure}

The polarization-limiting radius is a rough indication for the distance from the source at which the polarization of light is not affected by the birefringence anymore. In Fig.~\ref{fig:pol_rad}, the energy of the photon at which $r_p$ is equal to $r_I$ is plotted against the spin of the black hole (solid red line). The dotted line represents the ISCO (right $y-$axis). This means that, for rapidly spinning black holes, the effect of QED will be important around a photon energy of 10 keV or lower, while for slowly spinning black holes, QED will affect the polarization only above 10-20~keV. However, if the magnetic field strength is higher (or lower), the energy at which QED becomes important decreases (or increases) as the inverse square of the magnetic field strength. The effect of vacuum birefringence, if properly modeled, can therefore provide an indication on the strength of the magnetic field that threads the accretion disk. It is worth noticing that this result is dependent on the spin of the black hole but not on the mass, so it stands for both stellar-mass and supermassive black holes. The polarization-limiting radius estimate does not take into account light bending, which causes the photon's path in the strong magnetic field region to be longer due to the gravitational pull of the hole. For this reason, photons at energies lower than the one plotted in Fig.~\ref{fig:pol_rad} could still be affected by the vacuum birefringence, depending on their angular momentum (see Sec.~\ref{sec:case}).

\section{Competition with the plasma birefringence}
The possible presence of a corona above the inner regions of the disk introduces the possibility of a competing Faraday rotation due to the plasma birefringence. The effects of plasma birefringence for black hole accretion disks were studied in detail in a paper by \citet{2009ApJ...703..569D} and comprise of a reduction of the photons linear polarization in a range of energy that depends on the strength of the magnetic field, on the energy of the photons and on the distance to the black hole of the emission region. In this section, we estimate the photon energy above which the vacuum birefringence dominates over the plasma. If we write the two photon polarization modes as
\begin{subequations}
\begin{align}
|e_1 \rangle &= \cos\psi |a \rangle + i \sin\psi |b\rangle \\
|e_2\rangle &= \sin\psi |a\rangle - i \cos\psi |b\rangle
\end{align}
\end{subequations}
where
\begin{equation}
|a\rangle = \left(
\begin{array}{c}
-\sin\theta \\
0\\
\cos\theta
\end{array}
\right), \quad |b\rangle = \left(
\begin{array}{c}
0\\
1\\
0
\end{array}
\right),
\end{equation}
in the cold plasma limit we obtain
\begin{equation}
b = \frac{1}{\tan 2\psi} \simeq \frac{\omega_B}{\omega} \left[ 1 + V\frac{\omega^2-\omega_B^2}{\omega_B^2}\right] \frac{\sin^2\theta}{2\cos\theta}
\label{eq:bequation}
\end{equation}
where $\omega_B = eB/m_e c^2$ is the cyclotron frequency,
\begin{equation}
V = \frac{\alpha_{\rm{QED}}}{15\pi}\left(\frac{B}{B_{\rm{QED}}}\right)^2\left(\frac{\omega}{\omega_p}\right)^2
\end{equation}
measures the influence of the virtual $e^+$ $e^-$ pairs in the strong magnetic field relative to the real electrons of the plasma and $\omega_p$ is the plasma frequency \cite{1985ApJ...298..147M}.

For an accretion disk in the keV range, we are in the limit for which $\omega \gg \omega_B$. If $b$ goes to zero, the polarization becomes circular, and without the presence of QED, the Faraday rotation induced by the plasma would destroy the linear polarization, as in that case $b\simeq\omega_B/\omega \ll 1$. The limit for which the QED and the plasma effects are similar is for $b\sim 1$. Since $\omega_B/\omega \ll 1$, in order for $b$ to be about 1, $V\omega^2/\omega^2_B$ needs to be much greater than 1, so we can neglect the first term in the brackets of Eq.~(\ref{eq:bequation}) and then obtain
\begin{equation}
b \simeq \frac{\alpha_{\rm{QED}}}{15\pi}\left(\frac{B}{B_{\rm{QED}}}\right)^2\frac{\omega^3}{\omega^2_p\omega_B} = \frac{eBE^3}{60\pi^2 n_e m^2_e\hbar^2c^6}
\end{equation}
where $E$ is the energy of the photon and $n_e$ is the number density of electrons.

If we assume the optical depth over a distance comparable to the ISCO to be low:
\begin{equation}
\tau = n_e \sigma_T r_I \simeq 0.2
\end{equation}
where $\sigma_T$ is the Thomson cross section, 
we obtain that $b\sim 1$ for
\begin{equation}
E = 2.11-2.43 \; \mathrm{keV} \left(\frac{M}{10\,\mathrm{M}_\odot}\right)^{-\frac{1}{6}} \left(\frac{\tau}{0.2}\right)^{\frac{1}{3}}\left(\frac{1+X}{2}\right)^{\frac{1}{6}}
\end{equation}
where the first value is for $a_\star = 0$ and the second value is for $a_\star = 1$. Because $b$ scales as $E^3$, at higher energies the plasma birefringence does not destroy the linear polarization of the photons thanks to the predominance of QED which renders the propagation modes approximately linear.  The energy at which QED begins to dominate scales slowly with the assumed magnetic field strength, in fact as $B^{-1/3}$.

\section{Depolarization in the disk plane}
\label{sec:case}

To calculate in detail the effect of QED, as a first step we evolve the polarization of a photon traveling along a geodesic from the ISCO to the observer just above the plane of the accretion disk. We want to calculate the amount of depolarization caused by a partially ordered field in the disk. Recent measurements of the radio polarization around Sagittarius A$^*$ provided evidence for partially ordered magnetic fields near the event horizon \cite{2015Sci...350.1242J}. It is reasonable to assume the field to be organized on some length-scale that depends on the distance to the hole and that reflects the competition between the magnetic field itself, that tends to be organized, and the shear of the disk, impeding the formation of big structures. Therefore, to model the field in the disk, we divide the disk into regions of constant magnetic-field direction, which is the usual condition applied to the field in the equatorial plane in magnetohydrodynamics calculations \cite{2015MNRAS.446L..61P}.

As a photon travels through a magnetized birefringent vacuum with difference in index of refraction $\Delta n$, the polarization direction rotates around the birefringent vector $\bm{\hat{\Omega}}$ as
\begin{equation}
	\frac{d\Theta}{d\tau} = \Delta n \frac{p \cdot u}{\hbar c}
\end{equation}
where $p$ is the four-momentum of the photon, $u$ is the four-velocity of the disk that anchors the field and $\tau$ is the proper time elapsed in the frame of the disk. We want to calculate the final depolarization of the photon, so we integrate along the geodesic
\begin{equation}
\label{eq:int}
	\Delta \Theta = \int \Delta n \frac{p \cdot u}{\hbar c} \left(\frac{dx^\mu}{dr}\right)u_\mu dr 
\end{equation}
to determine the total rotation of the polarization of a photon across the Poincar\'e sphere. 
The polarization of an individual photon will perform a random walk across the Poincar\'e sphere, and the total rotation of the polarization along the path is given by Eq.~(\ref{eq:int}), where the extremes of the integral are the ISCO and infinity. The direction of the individual step, is given by Eq.~(\ref{eq:pol}).

For simplicity, we limit ourselves to photons traveling near the plane of the disk. In the equatorial plane, the metric becomes
\begin{subequations}
\begin{align}
	ds^2 &= g_{\mathrm{tt}}dt^2 + 2g_{\mathrm{t}\phi}drd\phi + g_{\phi\phi}d\phi^2 + g_{\mathrm{rr}}dr^2 \\
    g_{\mathrm{tt}} &= -1 + 2M/r \\
    g_{\mathrm{t}\phi} &= -2Ma/r \\
    g_{\phi\phi} &= r^2(1 + a^2/r^2 +2Ma^2/r^3) = r^2\mathcal{A} \\
    g_{\mathrm{rr}} &= (1 - 2M/r + a^2/r^2)^{-1} = \mathcal{D}^{-1}
\end{align}
\end{subequations}
The four-velocity of an observer rotating with the disk can be easily obtained remembering that
\begin{equation}
u^\phi = \frac{d\phi}{d\tau} = \omega u^{\mathrm{t}}
\end{equation}
From its definition, $g_{\mu\nu}u^\mu u^\nu = -1 $,
we obtain
\begin{subequations}
\begin{align}
	u^{\mathrm{r}} &=0 \\
	u^{\mathrm{t}} &= \sqrt{\frac{-1}{g_{\mathrm{tt}} + 2g_{\mathrm{t}\phi}\omega + g_{\phi\phi}\omega^2}} = \mathcal{BC}^{-\frac{1}{2}} \\
    u^\phi &= \frac{d\phi}{d\tau} = \omega u^{\mathrm{t}} = \omega \mathcal{BC}^{-\frac{1}{2}}
\end{align}
\end{subequations}
($u^{\mathrm{r}} =0$ because we are in the local orbiting frame), and
\begin{subequations}
\begin{align}
	u_{\mathrm{t}} &= (g_{\mathrm{tt}} +g_{\mathrm{t}\phi}\omega)u^{\mathrm{t}} = -\mathcal{GC}^{-\frac{1}{2}} \\
    u_\phi &= (g_{\phi\phi}\omega +g_{\mathrm{t}\phi})u^{\phi} = \sqrt{Mr} \mathcal{FC}^{-\frac{1}{2}}
\end{align}
\end{subequations}

In order to study the path of the photon along its geodesic, it is useful to calculate quantities that do not change along the path. From the dot-product of the four-momentum of the photon and two of the Killing vectors of the metric, we find two quantities that remain constant along the geodesics: the energy and angular momentum of the photon
\begin{subequations}
% * <heyl@phas.ubc.ca> 2017-10-12T03:49:54.232Z:
% 
% The photon doesn't have a four velocity, so I replaced it with the four-momentum of the photon which you have already defined.
% 
% ^.
\begin{align}
	\label{eq:E}
	E &= -\xi_{\mathrm{t}} \cdot p = - (g_{\mathrm{tt}}p^{\mathrm{t}} + g_{\mathrm{t}\phi}p^\phi) \\
    \label{eq:L}
	L &= \xi_{\phi} \cdot p = g_{\phi\phi}p^{\phi} + g_{\mathrm{t}\phi}p^{\mathrm{t}}
\end{align}
\end{subequations}
We call the specific angular momentum $L/E = l$.
We analyze three cases: a photon coming from the ISCO with zero angular momentum ($l = 0$), a photon initially rotating with the disk (maximum prograde $l_+$) and a photon initially going against the rotation of the disk (maximum retrograde $l_-$).

\subsection{Zero angular momentum photons}

If $l = 0$, from Eqs.~(\ref{eq:E}) and~(\ref{eq:L}), we obtain, for the photon,
\begin{equation}
	\frac{d\phi}{dt}=-\frac{g_{\mathrm{t}\phi}}{g_{\phi\phi}} = 2\frac{Ma}{r^3}\mathcal{A}^{-1}
\end{equation}
From the null-geodesic condition $ds^2=0$, we find:
\begin{equation}
		\frac{dt}{dr}= \sqrt{\frac{g_{\mathrm{rr}}}{\frac{g_{\mathrm{t}\phi}^2}{g_{\phi\phi}} - g_{\mathrm{tt}}}} = \mathcal{A}^{\frac{1}{2}}\mathcal{D}^{-1}
\end{equation}
We can then write the second part of Eq.~(\ref{eq:int}) as
\begin{align}
	\left(\frac{dx^\mu}{dr}\right)u_\mu &=\frac{d\phi}{dr}u_\phi + \frac{dt}{dr}u_{\mathrm{t}} \nonumber \\
    &= - (\mathcal{AC})^{-\frac{1}{2}}\mathcal{B}.
\end{align}
The first part becomes:
\begin{equation}
p\cdot u = E(-u^{\mathrm{t}} + b u^\phi) = -Eu^{\mathrm{t}}
\end{equation}
Using $\Delta n$ from Eq.~(\ref{eq:Omega}), and changing the integration variable to a dimensionless one ($r_\star = r/M$), Eq.~(\ref{eq:int}) becomes
\begin{equation}
\label{eq:zero}
\Delta\Theta =  E K \int \sin^2\theta r_\star^{-\frac{3}{2}}\mathcal{A}^{-\frac{1}{2}}\mathcal{B}^2\mathcal{N}(\mathcal{DC})^{-1} dr_\star
\end{equation}
where $K = m_p/[15\pi m_e^2c^2(1+X)]$.

\subsection{Maximum prograde and retrograde angular momentum photons}

From Eqs.~(\ref{eq:E}) and~(\ref{eq:L}), we obtain, for the photon,
\begin{equation}
	\label{eq:44}
	\frac{d\phi}{dt}=- \frac{l g_{\mathrm{tt}} + g_{\mathrm{t}\phi}}{g_{\phi\phi} + l g_{\mathrm{t}\phi}}
\end{equation}
By imposing $dr^2=0$, at the point of emission (the ISCO), we obtain the values for the maximum prograde specific angular momentum ($l_+$) of a photon rotating with the disk and the maximum retrograde specific angular momentum ($l_-$) for a photon in retrograde motion:
\begin{align}
	l_{\pm} &= \frac{g_{\mathrm{t}\phi} \pm 	\sqrt{g_{\mathrm{t}\phi}^2 - 	g_{\phi\phi}g_{\mathrm{tt}}}}{-g_{\mathrm{tt}}} \\
	&= r \left( \frac{-2 a_\star/r_\star^2 \pm \mathcal{D}^{1/2}}{1 - 2/r_\star} \right) \nonumber
\end{align}
Since $l$ is a constant along the geodesic, we calculate $l_\pm$ at the ISCO. Depending on the spin of the black hole, however, the ISCO can be inside the retrograde photon orbit (the prograde photon orbit is always inside the ISCO). In this case, we calculate $l_-$ at the retrograde photon orbit.

Employing Eq.~(\ref{eq:44}) in the null-geodesic condition, we find the path of the photon:
\begin{subequations}
\begin{align}
\frac{dt}{dr} &= \frac{g_{\phi\phi} + l g_{\mathrm{t}\phi}}{r\mathcal{D}(l^2g_{\mathrm{tt}}+2lg_{\mathrm{t}\phi}+g_{\phi\phi})^{1/2}} \\
\frac{d\phi}{dr} &= \frac{- (lg_{\mathrm{tt}} +g_{\mathrm{t}\phi})}{r\mathcal{D}(l^2g_{\mathrm{tt}}+2lg_{\mathrm{t}\phi}+g_{\phi\phi})^{1/2}}
\end{align}
\end{subequations}
Defining a dimensionless angular momentum as $l_\star = l/M$, we can then write the second part of Eq.~(\ref{eq:int}) as
\begin{align}
	\left(\frac{dx^\mu}{dr}\right)u_\mu &=\frac{d\phi}{dr}u_\phi + \frac{dt}{dr}u_{\mathrm{t}} \nonumber \\
    &= \frac{l\sqrt{Mr} - r^2 \mathcal{B}}{r\mathcal{C}^{1/2}(l^2g_{\mathrm{tt}}+2lg_{\mathrm{t}\phi}+g_{\phi\phi})^{1/2}} \nonumber\\
    &= \frac{l_\star/r_\star^{3/2} - \mathcal{B}}{\mathcal{C}^{1/2}(l_\star^2g_{\mathrm{tt}}/r_\star^2-4l_\star a_\star/r_\star^3 + \mathcal{A})^{1/2}}
\end{align}
The first part becomes
\begin{align}
p\cdot u &= E(-u^{\mathrm{t}} + l u^\phi) \nonumber \\
&= E\mathcal{C}^{-1/2} (l_\star/r_\star^{3/2} - \mathcal{B})
 \end{align}
We can then rewrite Eq.~(\ref{eq:int}) as
\begin{widetext}
\begin{equation}
\label{eq:maxmin}
\Delta\Theta =  E K \int \sin^2\theta r_\star^{-\frac{3}{2}}\frac{\mathcal{N}}{\mathcal{D}\mathcal{C}} \frac{(l_\star/r_\star^{3/2} - \mathcal{B})^2}{(l_\star^2g_{\mathrm{tt}}/r_\star^2-4l_\star a_\star/r_\star^3 + \mathcal{A})^{1/2}} dr_\star
\end{equation}
\end{widetext}
where $K = m_p/[15\pi m_e^2c^2(1+X)]$ is the same as in the previous section.
% * <heyl@phas.ubc.ca> 2017-10-12T04:17:23.268Z:
% 
% I worked on everything up to here ... because you have just started the results section.  It is looking really good!
% 
% ^.
\subsection{Results}

As explained at the beginning of this section, we assume a partially organized field in the disk, as we divide the disk into region of fixed magnetic field direction. In order to model the magnetic field of the disk, we need to make an assumption on the length-scale over which the magnetic field is organized, \textit{i.e.} the length-scale of the annular regions. As we expect the length-scale to be related to both the size of the hole itself and the distance to the hole, we first divide the disk into 5 regions, each twice as large as the previous one: from the ISCO to twice the ISCO, to 4 times the ISCO, to 8 times the ISCO, to 16 times the ISCO, to infinity. For simplicity, we call this configuration the ``2-fold configuration". Equations~(\ref{eq:zero}) and~(\ref{eq:maxmin}) allow us to calculate the path that the polarization of a photon takes across the Poincar\'e sphere in each region. To calculate the direction of the step, we rotate \textbf{s} around $\hat{\mathbf{\Omega}}$ [Eq.~(\ref{eq:pol})]. In each region, we take the angle between the magnetic field and the photon, $\theta$, and the angle between \textbf{s} and $\hat{\mathbf{\Omega}}$ as random.

\begin{figure*}
  \includegraphics[width=0.321\textwidth]{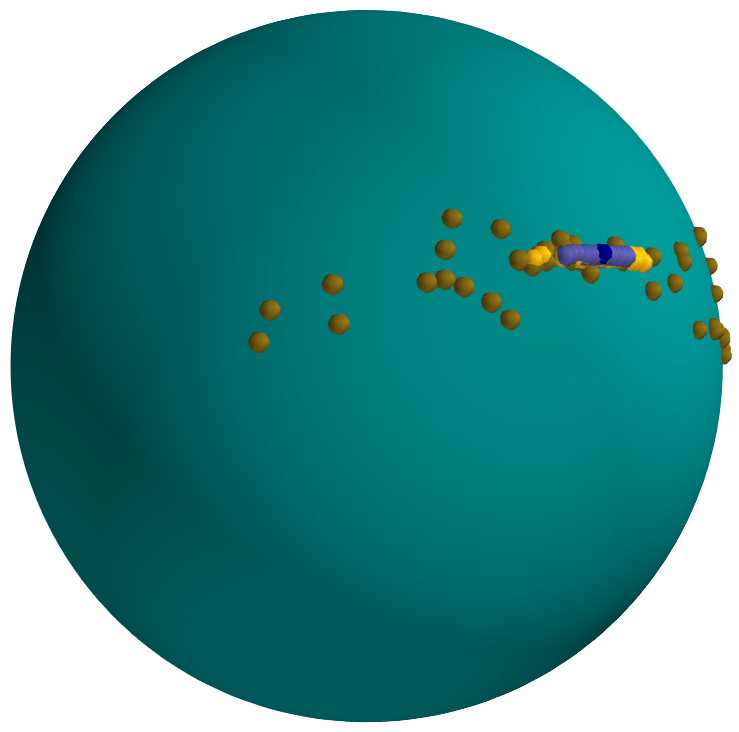}
  \includegraphics[width=0.32\textwidth]{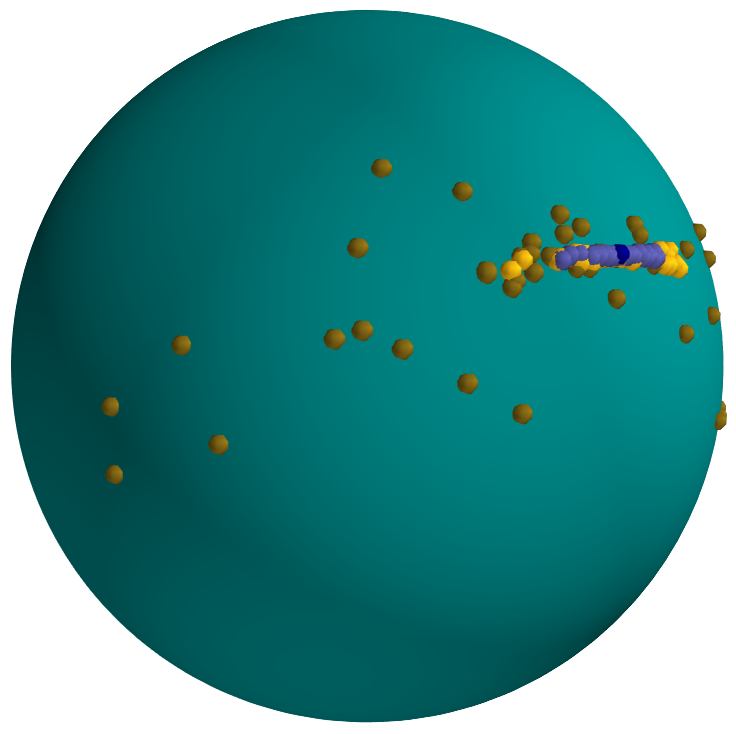}
  \includegraphics[width=0.318\textwidth]{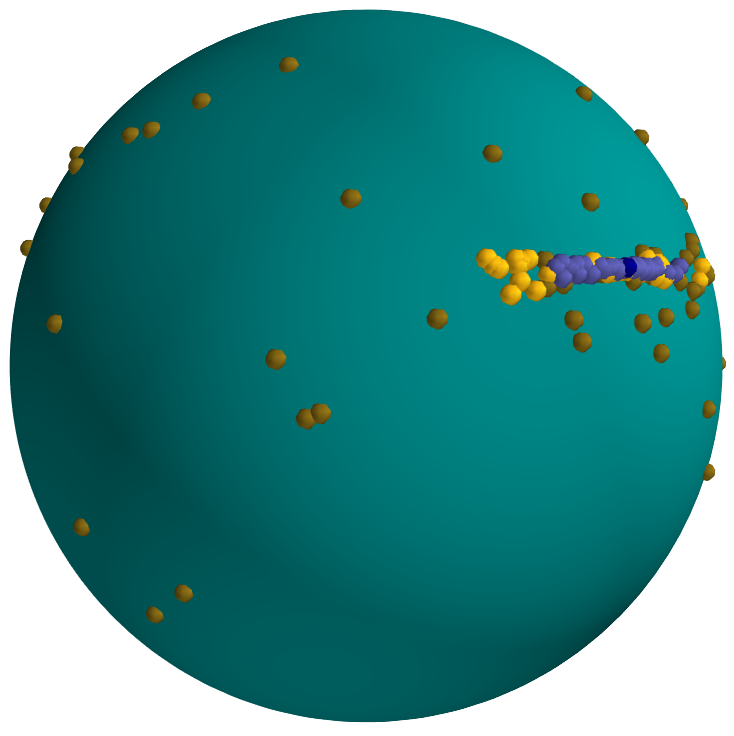}
\caption{Monte-Carlo simulation of the depolarization of radiation from a black hole with $a = 0.84$ (as NGC 1365) for three photon energies (as measure by a distant observer): 3 keV (left), 5 keV (middle) and 7 keV (left). Polarization is represented on the Poincar\'e sphere: the dots represent the end-point of the polarization vector. The initial polarization vector is indicated by a dark blue dot. The violet dots are photons that receive a large blue shift (90\% of $l_+$) , the yellow dots are zero-angular-momentum photons and the copper receive a large red shift (90\% of $l_-$) on their way from the ISCO to us.}
\label{fig:spheres}
\end{figure*}

In order to visualize the depolarization effect of the partially ordered field on the single photon, we first performed a Monte Carlo simulation for 60 photons, calculating the evolution of their polarization from the ISCO to infinity. Each photon is emitted with the same angular momentum and the same energy at infinity from the ISCO of a black hole rotating with $a_\star=0.84$ (as the AGN NGC 1365 \cite{2013Natur.494..449R}). We repeated the same calculation for photons traveling with zero, 90\% of the maximum prograde and 90\% of the maximum retrograde specific angular momentums and for three different energies: 3, 5 and 7 keV (at infinity). The results are shown in Fig.~\ref{fig:spheres}.
Figure~\ref{fig:spheres} depicts a solid Poincar\'e sphere, in which the dots represent the end-point of the polarization vectors. The dark blue dot indicates the initial polarization, which is the same for every photon. Without the QED effect, the polarization would be frozen at the emission and the final polarization at infinity would be the same for all photons: still the dark blue dot. The other dots indicate the final polarization of the photons, calculated within QED. The yellow dots indicate the end-point of the polarization vector for the zero angular momentum photons; the violet dots correspond to the photons that receive a large blue shift ($l_+$ photons) and the copper dots represent the photons that receive a large red shift ($l_-$ photons). We can immediately see that the final polarization is different from the initial one for all the photons, with a much bigger effect for red-shifted photons and for high-energy photons.

\begin{figure*}
  \includegraphics[width=0.49\textwidth]{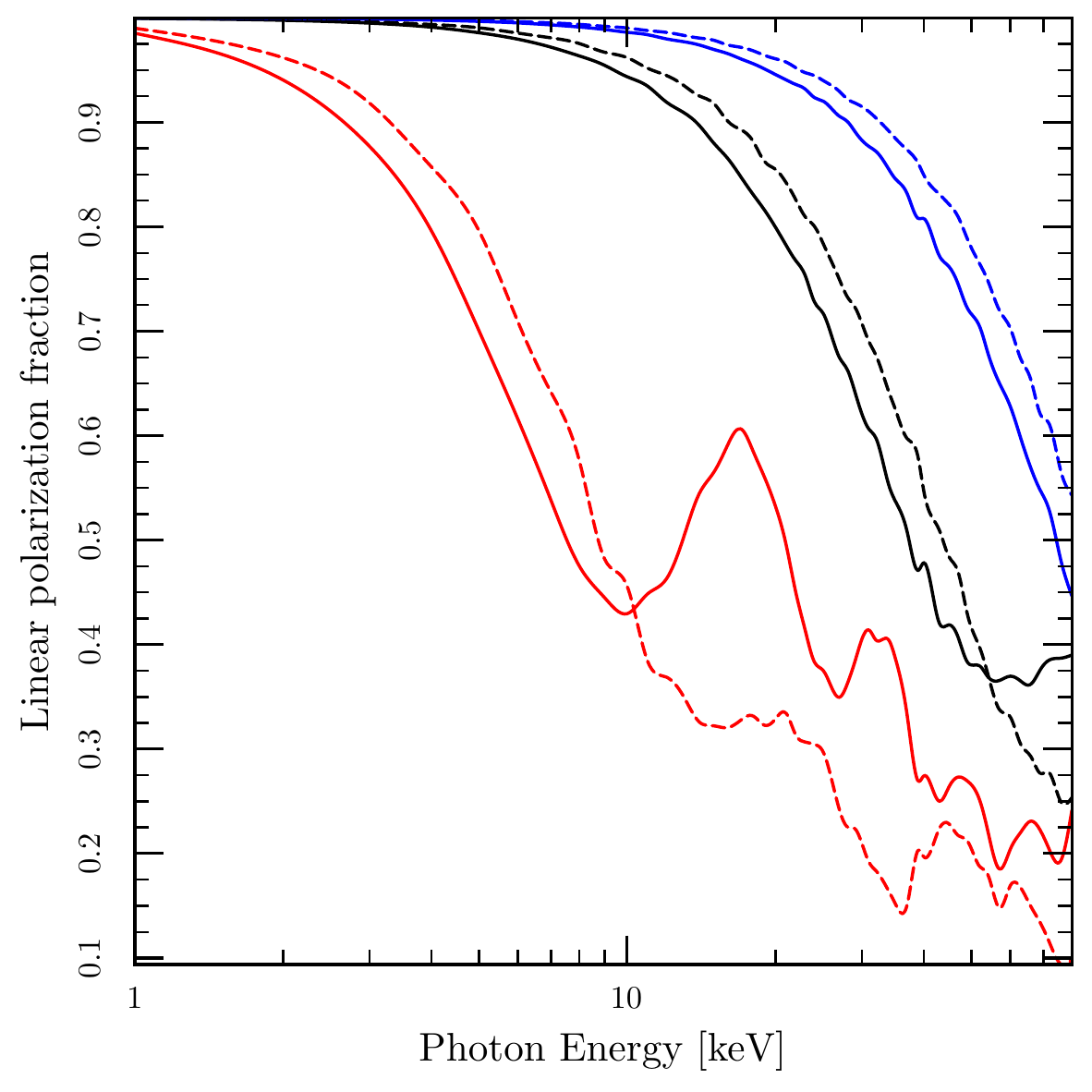}
  \includegraphics[width=0.49\textwidth]{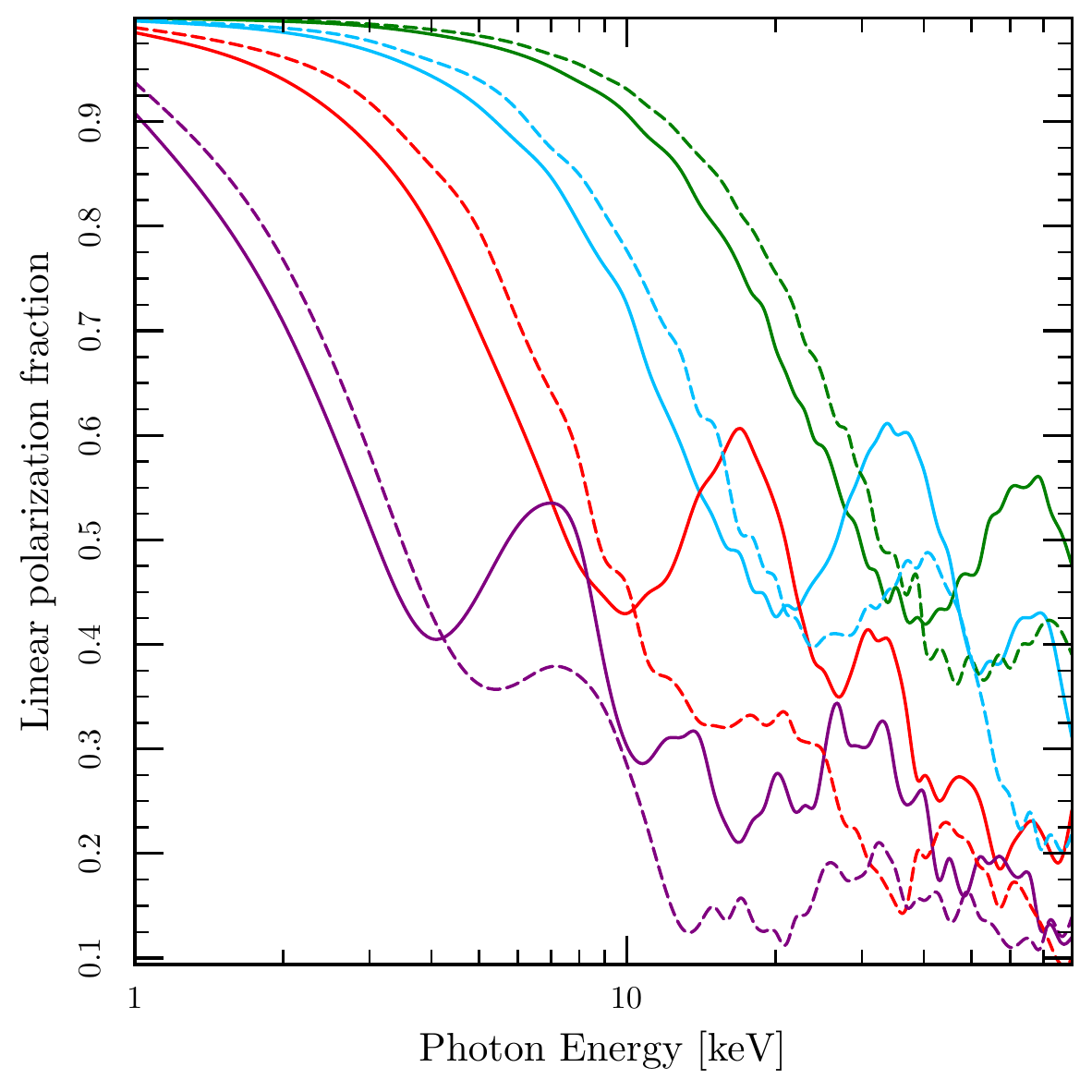}
\caption{Final polarization fraction vs.\ photon energy calculated in the 2-fold configuration (solid lines) and in the 1.5-fold configuration (dashed lines). Left plot, left to right: maximum retrograde (90\% $l_-$) angular momentum photons (red), zero angular momentum photons (black) and maximum prograde (90\% $l_+$) angular momentum photons (blue), coming from the ISCO of a black hole with $a_\star=0.9$. Right plot: 90\% $l_-$ photons for, left to right, $a_\star=0.99$ (purple), 0.9 (red), 0.7 (light blue) and 0.5 (green).}
\label{fig:plots}
\end{figure*}

We then performed the same Monte-Carlo simulation, this time with 6,000 photons, for different energies in the upcoming polarimeters' bands, 1 to 80~keV (at the observer), and for four values of $a_\star$: 0.5, 0.7, 0.9 and 0.99. In order to check in which way our assumption on the structure of the magnetic field affects our results, we performed the same calculations also for a disk in which the regions of constant magnetic-field direction are, each, 1.5 times as large as the previous one: from the ISCO to 1.5 times the ISCO, to 2.3 times the ISCO, to 5.1 times the ISCO, to 7.6 times the ISCO, to 11 times the ISCO, to 17 times the ISCO, to infinity. For simplicity, we call this configuration the 1.5-fold configuration. The results are shown in Fig.~\ref{fig:plots}. Both plots show the polarization fraction obtained as an average of the final linear polarization of all the 6,000 photons against the photon energy. Results are shown for both the 2-fold configuration (solid lines) and the 1.5-fold configuration (dashed lines). The left plot shows the final polarization fraction of the zero angular momentum photons (black lines), the blue-shifted photons (blue lines) and the red-shifted photons (red lines) for a black hole rotating with $a_\star=0.9$. The right plot shows the polarization fraction of red-shifted photons for four different $a_\star$: 0.5 (green lines), 0.7 (light blue lines), 0.9 (red lines) and 0.99 (purple lines).

All our results are independent of the black hole mass.

\section{Conclusions}
As we can see from the left panel of Fig.~\ref{fig:plots}, the effect of the vacuum birefringence on the polarization of x-ray photons is important, especially for retrograde photons at the high end on the upcoming polarimeters' range, where the observed polarization can be reduced up to 90\% for a black hole rotating at 90\% the critical spin. There are two reason why retrograde photons are more affected by vacuum birefringence: they are the ones that undergo more orbits around the hole and the energy at which they are emitted is higher. The effect of QED is larger for rapidly rotating black holes, for which the magnetic field at the ISCO is higher and photons perform many rotations around the hole before leaving the magnetosphere (see the right panel of Fig.~\ref{fig:plots}).

The dashed lines plot in both panels of Fig.~\ref{fig:plots} show the averaged final polarization fraction for photons traveling through a different field configuration (the 1.5-fold configuration). In general, if magnetic loops are smaller, the depolarization effect is reduced linearly with the size of the loops: in our example, the dashed lines fall on top of the solid lines if we rescale them by 2/1.5. However, the solid lines show peaks that are not present in the dashed lines. For example, for a hole rotating with spin $a_\star = 0.99$  in the 2-fold configuration (purple solid line, right panel) the polarization fraction peaks at 7 keV and then again at 14 keV, at 21 keV and so on. These peaks are due to the fact that at those energies the integral in Eq.~(\ref{eq:maxmin}) reaches, in the first zone of the disk, an average value of $\pi$, and therefore, the polarization vector remains closer to the $S_1-S_2$ plane. In the 1.5-fold configuration this does not happen because the first region is smaller and the second region has a bigger effect on the final polarization, washing out the peaks. Ideally, the presence of features in the polarization spectrum like the peaks shown for the 2-fold configuration could provide hints on the structure of the magnetic field in the disk.

Our analysis is restricted to nearly edge-on observations of black-hole accretion disks, in which the photons that reach the observer are traveling very close to the disk through the magnetosphere. Further studies are needed for modeling photons coming out of the plane of the disk and traveling through a more organized magnetic field structure. In this case, the effect of QED could be opposite to the one due to a chaotic field in the disk. Indeed, the presence of an organized field could increase the linear polarization of photons instead of destroying it. 

All the results in this paper are obtained for the minimum magnetic field needed for accretion to operate in the disk. A larger magnetic field would cause a higher depolarization at all energies. The observation of the x-ray polarization from black-hole accretion disks would provide, if properly modeled including QED, a measurement of the strength of the magnetic field in the disk and therefore a validity check for the theories of astrophysical accretion.

{\noindent \bf Acknowledgements}

We thank the anonymous referee for many useful suggestions that improved the paper significantly.   We used the NASA ADS service and arXiv.org. This work was supported by the Natural Sciences and Engineering Research Council of Canada (FundRef ID \url{http://dx.doi.org/10.13039/501100000038}, Discovery Grant), the Canadian Foundation for Innovation (FundRef ID \url{http://dx.doi.org/10.13039/501100000196}) and the British Columbia Knowledge Development Fund (FundRef ID \url{http://dx.doi.org/10.13039/501100007711}).  I.C.\ was supported by a Four-Year Fellowship from the Faculty of Graduate Studies at University of British Columbia (FundRef ID \url{http://dx.doi.org/10.13039/501100000187}).

\bibliography{blackholesQED}

\end{document}